\documentclass[reprint,
 prl,amsmath,amssymb,
 aps]{revtex4-2}

\usepackage{graphicx}
\usepackage{mathtools}
\usepackage{dcolumn}
\usepackage{bm}
\usepackage{amsthm}
\usepackage{amsfonts}
\usepackage{bbm}
\usepackage{comment}
\usepackage{hyperref}
\usepackage[mathlines]{lineno}

\begin{document}

\title{Skin effect and winding number in disordered non-Hermitian systems}

\author{Jahan Claes}
\author{Taylor L. Hughes}
\affiliation{Department of Physics and Institute for Condensed Matter Theory$,$\\
 University of Illinois at Urbana-Champaign$,$ Illinois 61801$,$ USA}

\begin{abstract}
Unlike their Hermitian counterparts, non-Hermitian (NH) systems may display an exponential sensitivity to boundary conditions and an extensive number of edge-localized states in systems with open boundaries, a phenomena dubbed the ``non-Hermitian skin effect." The NH skin effect is one of the primary challenges to defining a topological theory of NH Hamiltonians, as the sensitivity to boundary conditions invalidates the traditional bulk-boundary correspondence. The NH skin effect has recently been connected to the winding number, a topological invariant unique to NH systems. In this paper, we extend the definition of the winding number to disordered NH systems by generalizing established results on disordered Hermitian topological insulators. Our real-space winding number is self-averaging, continuous as a function of the parameters in the problem, and remains quantized even in the presence of strong disorder. We verify that our real-space formula still predicts the NH skin effect, allowing for the possibility of predicting and observing the NH skin effect in strongly disordered NH systems. As an application we apply our results to predict a NH Anderson skin effect where a skin effect is developed as disorder is added to a clean system, and to explain recent results in optical funnels.
\end{abstract}

\maketitle

\section{Introduction}

The topological classification of Hermitian Hamiltonians\cite{qi2011topological,hasan2010colloquium,bernevig2013topological,schnyder2008classification,qi2008,kitaev2009periodic,ryu2010topological,chiu2016classification} applies to non-interacting Hamiltonians in equivalence classes based on  the ten internal Altland-Zirnbauer (AZ) symmetry classes\cite{altland1997nonstandard}. The topological classifications for all of these symmetries are known for all spatial dimensions, and explicit forms for the corresponding topological invariants have been constructed\cite{schnyder2008classification,qi2008,kitaev2009periodic,ryu2010topological,chiu2016classification}. A notable feature of the Hermitian topological classification is the celebrated \emph{bulk-boundary correspondence}, in which topological invariants of the bulk system predict anomalous states on the boundary\cite{qi2011topological,hasan2010colloquium,bernevig2013topological}.

There has been recent interest in the topological properties of non-Hermitian (NH) Hamiltonians\cite{torres2019perspective,hu2011,alvarez2018topological,bergholtz2019exceptional,el2018non,gong2018topological,yao2018edge,leykam2017edge,esaki2011edge,lee2016anomalous,yao2018non,lieu2018topological,ozawa2019topological,weimann2017topologically,harari2018topological,malzard2015topologically}. NH Hamiltonians provide effective models for quantum systems with gain and loss\cite{rotter2009non}, and can be realized in atomic\cite{xu2017weyl,lee2014heralded,li2019observation,yamamoto2019theory,nakagawa2018non}, optical\cite{malzard2015topologically,makris2008beam,schomerus2013topologically,ozawa2019topological,xiao2017observation,zhen2015spawning,malzard2018bulk,harari2018topological,cerjan2019experimental,chen2017exceptional,xiao2020non,weimann2017topologically,weidemann2020topological}, electronic\cite{hofmann2020reciprocal,helbig2020generalized,jiang2019interplay} and mechanical\cite{ghatak2019observation,brandenbourger2019non,zhou2020non,schomerus2020nonreciprocal,scheibner2020non,yoshida2019exceptional} systems. Like their Hermitian counterparts, NH Hamiltonians can display protected anomalous boundary states\cite{alvarez2018topological,bergholtz2019exceptional,lee2016anomalous}. In addition, they display topological phenomena unique to NH systems, such as exceptional points\cite{kozii2017non,lee2016anomalous}, half-integer winding\cite{lee2016anomalous}, stable 2D semimetallic phases\cite{yoshida2019symmetry,kozii2017non,kawabata2019classification,budich2019symmetry}, and Weyl exceptional rings\cite{xu2017weyl,cerjan2019experimental}. All of these phenomena are tied to a richer set of symmetry classes beyond the ten AZ classes\cite{kawabata2019symmetry,zhou2019periodic,kawabata2019classification,budich2019symmetry}.

A primary challenge for developing a theory for the topological phenomena of NH Hamiltonians is the NH skin effect, in which systems may display remarkably different eigenspectra and eigenstates in periodic vs open boundary conditions (PBC or OBC) \cite{alvarez2018non,lee2019anatomy,yao2018edge,xiong2018does,kunst2018biorthogonal}. In particular, PBC and OBC systems might become gapless at different points in their phase diagrams\cite{xiong2018does,yokomizo2019non,kunst2018biorthogonal,yao2018edge, yao2018non}, and topologically protected edge states may be hidden in an extensive number of edge-localized eigenstates. In the presence of the NH skin effect, topological invariants calculated using PBC may not predict properties of the OBC system. One way to address this issue is by computing the so-called generalized Brillouin zone\cite{yao2018non,yao2018edge,yokomizo2019non}, which can be done only for simplified models, or by using real-space invariants that directly predict OBC properties\cite{kunst2018biorthogonal,luo2019non,song2019non,zhang2020non}. However, in general, a precise understanding of the NH skin effect is necessary to develop a bulk-boundary correspondence for NH systems\cite{jin2019bulk}. In addition, the NH skin effect is also interesting in its own right, as systems with the NH skin effect exhibit exponentially large responses to perturbations\cite{ghatak2019observation,schomerus2020nonreciprocal}, and optical systems exhibiting the NH skin effect have recently been demonstrated to funnel light for high-performance optical sensors\cite{weidemann2020topological}.

While understanding the NH skin effect and its relationship to the bulk-boundary correspondence is still an outstanding question in general dimensions, in 1D it has recently been shown that the NH skin effect is determined by the winding number around a complex energy $E$: $w(E)$\cite{borgnia2020non,okuma2020topological,zhang2019correspondence}. The winding number is a topological invariant unique to NH systems\cite{gong2018topological,kawabata2019classification} and,  roughly speaking, regions of the complex energy plane with nonzero $w(E)$ under PBC have dramatically different spectra under OBC. Furthermore, the sign of $w(E)$ determines the edge on which an extensive number of states are localized. Thus, determining the winding number $w(E)$ provides a criterion for the presence or absence of a NH skin effect in the neighborhood around $E$.

In this work, we generalize the winding number $w(E)$ to disordered NH systems. We accomplish this by mapping the the NH problem to a disordered Hermitian system, for which previous results have been established. As a result we propose a real-space formula for $w(E)$ that is self-averaging, continuous as a function of the parameters in the problem, and remains quantized even in the presence of strong disorder. We verify the stability of our invariant in simple models of NH systems with nontrivial winding. We also demonstrate that our invariant $w(E)$ determines the existence of the NH skin effect and, when non-vanishing, also predicts the existence of an entire band of delocalized states in our system. Indeed, the non-zero winding essentially protects these states from localization, which is in striking contrast to disordered 1D Hermitian systems which are always localized (unless tuned to a critical point)\cite{song2014aiii,mondragon2014topological}.

We note that the NH skin effect has already been observed in  optical\cite{xiao2020non}, electronic\cite{hofmann2020reciprocal,helbig2020generalized}, and mechanical\cite{ghatak2019observation,brandenbourger2019non,zhou2020non} systems, so our results can be experimentally realized in multiple physical contexts. In addition, our results have positive implications for the stability of the recently demonstrated optical funnel\cite{weidemann2020topological} to disorder. Finally, since our method holds at arbitrary disorder strength, our results predict the possibility of observing a ``NH Anderson skin effect," an NH analogue of the topological Anderson insulator\cite{song2014aiii,li2009topological,groth2009theory,mondragon2014topological,meier2018observation,luo2019non,zhang2020non}, in which the skin effect is induced entirely by disorder from a clean system without a skin effect.

\section{The non-Hermitian winding number}

\begin{figure}
    \centering
    \includegraphics[width=\columnwidth]{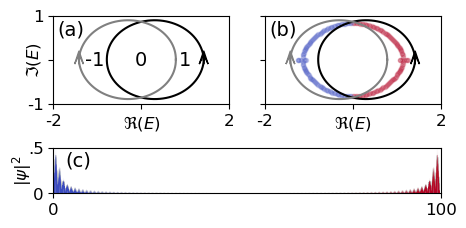}.
    \caption{An example of a NH band structure with $+1$ and $-1$ winding. (a) The PBC bands in the complex plane, with the winding numbers marked. (b) The same diagram, with the OBC spectrum overlaid. Red denotes states in a region with $w(E)>0$, blue denotes states in a region with $w(E)<0$. We see that the regions with positive/negative winding collapse onto lines. (c) The density $|\psi^n(x)|^2$ of all OBC eigenstates $\{\psi^n\}$. We see that eigenstates in regions with negative winding (blue) are localized at the left edge, while eigenstates in regions with positive winding (red) are localized at the right edge.}
    \label{fig:windingExample}
\end{figure}

A translationally invariant NH Hamiltonian can be decomposed by the usual Fourier transform $\hat{H}=\oplus_k\hat{H}_k$, where $\hat{H}_k=e^{ik\hat X}\hat H e^{-ik\hat X}$ and $\hat X$ is the position operator. As in the Hermitian case, the eigenstates of $\hat{H}$ can be written as  $|\psi_k^n\rangle=e^{ik\hat{X}}|u_k^n\rangle$, with $\hat{H}_k|u_k^n\rangle=E_k^n$. Unlike in the Hermitian case, however, the eigenvalues $E_k^n$ can in general be complex. Fig. \ref{fig:windingExample}(a) illustrates possible curves $E_k^n$ for a two band model. 
Because of the complex eigenvalues, we can distinguish two types of gaps in a NH spectrum. \emph{Point gaps} are values $E\in \mathbb C$ such that the bands $E_k^n$ never intersect $E$, while \emph{line gaps} are lines $\ell\subset\mathbb{C}$ such that the bands never intersect $\ell$\cite{gong2018topological,kawabata2019symmetry}. These gaps coincide for Hermitian systems, but are distinct for NH systems.

For any point gap $E$, we can define a topological invariant $w(E)$ that simply counts how many times the bands $E_k^n$ wind around $E$. This is a topological invariant because $w(E)$ cannot change unless the gap closes at the base value $E$. We can write a formula for the winding number\cite{gong2018topological}:
\begin{equation}
    w(E) = \frac{1}{2\pi i}\int_0^{2\pi} \partial_k \log\left( |\hat H_k-E|\right)\ dk.
    \label{eq:winding}
\end{equation}
This formula counts the number of times the determinant of $(\hat{H}_k-E)$ winds around the origin, which is equivalent to counting the total winding of the bands $E_k^n$. The winding numbers of different regions are indicated by the integers in Fig. \ref{fig:windingExample}(a).

It has recently been proven that the winding number predicts the NH skin effect\cite{borgnia2020non,okuma2020topological,zhang2019correspondence}. Concretely, when transitioning between PBC and OBC, the PBC eigenvalues enclosing a region having winding $w\neq 0$ collapse onto 1D arcs within the region. An example is shown in Fig. \ref{fig:windingExample}(b), where the  OBC eigenvalues appear in the interior of the regions having $w=\pm 1$. Moreover Refs. \cite{okuma2020topological,zhang2019correspondence} proved that the OBC eigenvalues located in the interior of regions having $w>0$ correspond to eigenstates localized at the right edge of the system, and the OBC eigenvalues located in the interior of regions where $w<0$ correspond to eigenstates localized at the left edge of the system, as is shown in Fig. \ref{fig:windingExample}(c). The number of eigenvalues in each arc is proportional to the system size, leading to an extensive number of states at the corresponding edge. The marked difference between OBC and PBC spectra and the extensive number of edge states always occur together, and collectively make up the NH skin effect\cite{yao2018edge}.

\section{The disordered non-Hermitian winding Number}

Ref. \onlinecite{gong2018topological} previously introduced a formula for the winding number in the presence of disorder. Adding a flux $\phi$ through the periodic system, they define the winding number to be
\begin{equation}
    w(E) = \frac{1}{2\pi i}\int_0^{2\pi} \partial_\phi \log\left( |\hat H(\phi)-E|\right)\ d\phi.
\end{equation}
This formula has been successfully applied to the a non-Hermitian version of the quasiperiodic Aubry-André-Harper model\cite{harper1955single,aubry1980analyticity} to predict topological localization transitions\cite{longhi2019topological,zeng2020winding}, mobility edges\cite{zeng2020winding}, and the NH skin effect\cite{jiang2019interplay}. However, for general models at strong disorder, this formula a few drawbacks. Most notably, it requires evaluating the expression at a large number of $\phi$ points to estimate the integral. In addition, it is not clear that this formula is self-averaging for large systems. Finally, it is not obvious that this formula is well-behaved at strong disorder when eigenvalues will generically exist near $E$ such that the phase of $|\hat{H}(\phi)-E|$ is sensitive to rounding errors.

In this paper, we instead define a real-space formula for the winding number $w(E)$ for disordered systems using techniques from non-commutative geometry. Our formula relies on mapping a NH Hamiltonian $\hat{H}$ to a doubled Hermitian Hamiltonian $\hat{\mathcal{H}} =\hat\sigma_+\otimes (\hat{H}-E)+\hat\sigma_-\otimes (H-E)^\dagger$ with chiral symmetry $\hat{S}=\hat{\sigma_z}\otimes\mathbbm{1}$\cite{gong2018topological}, and applying known results from non-commutative geometry to $\hat{\mathcal{H}}$ \cite{mondragon2014topological,song2014aiii,prodan2016non,prodan2016bulk}. Concretely, we define an operator $\hat{Q}$ by the polar decomposition $(\hat{H}-E)=\hat{Q}\hat{P}$, where $\hat Q$ is unitary and $\hat{P}$ is positive. Refs. \onlinecite{herviou2019defining,gong2018topological} have previously used $\hat Q$ to define topological properties of clean NH systems. In terms of $\hat{Q}$, we define
\begin{equation}
    w(E) = \mathcal{T}(\hat Q^\dagger [\hat Q,\hat X]),
    \label{eq:realWinding}
\end{equation}
where $\mathcal{T}$ is the trace per unit volume. This reduces to Eq. \ref{eq:winding} when the system is translationally invariant. In addition, $w(E)$ is quantized, self-averaging, continuous as a function of $E$ and parameters in the Hamiltonian, and only changes when there are states with diverging localization length $\Lambda(E)$ at $E$. Furthermore, a semi-infinite system has an eigenstate with eigenvalue $E$ localized at the boundary whenever $w(E)\neq 0$, which provides a justification of Eq. \ref{eq:realWinding} as the real-space generalization of the winding number (see Supplement\footnote{See Supplemental Material at [URL will be inserted by publisher] for a detailed derivation of our formula for $w(E)$ its properties.} for more details of the proof). Our formula for $w(E)$ is a real-space NH invariant in the spirit of Refs. \onlinecite{luo2019non,song2019non,zhang2020non}, although unlike those works our invariant has no Hermitian counterpart.

\section{Hatano-Nelson model with disorder}

To illustrate the properties of $w(E)$ formula, we consider a disordered Hatano-Nelson model \cite{hatano1996localization,hatano1997vortex,hatano1998non}:
\begin{equation}
    \hat{H} = \sum_i J_R^i \hat{c}_{i+1}^\dagger \hat{c}_i+J_L^i \hat{c}_{i}^\dagger \hat{c}_{i+1} +h^i \hat{c}_i^\dagger \hat{c}_i.
\end{equation}
This model describes a uniform chain with independent hoppings $J_L^i$ and $J_R^i$ in the left and right directions, and an on-site potential $h^i$ (which we take to be real). This model is non-Hermitian when $J_L^i\neq J_R^i$. To model disorder we choose the hopping parameters independently according to
\begin{equation}
    J_R^i = J_R +W_R\omega^i_R,\quad J_L^i = J_L +W_L\omega^i_L, \quad h^i=W\omega^i,
    \label{eq:disorderParameters}
\end{equation}
where $\omega_{(L/R)}^i\in [-.5,.5]$ are uniformly distributed random variables. 

\begin{figure}
    \centering
    \includegraphics[width=\columnwidth]{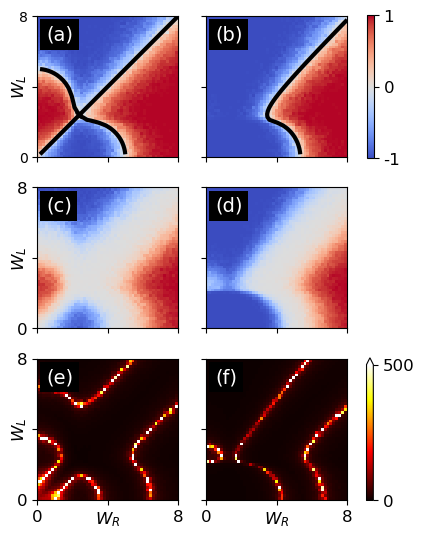}.
    \caption{(a) $w(0)$ as a function of $(W_R,W_L)$ for $(J_L,J_R,W)=(1,1,0)$. The black line denotes the points where $\Lambda(0)$ diverges. (b) Same plot for $(J_L,J_R,W)=(1,.5,0)$. (c,d) $w(E)$ as a function of $(W_R,W_L)$ for $(J_L,J_R,W)=(1,1,1)$ and $(J_L,J_R,W)=(1,.5,1)$. (e,f) Numerically computed $\Lambda(0)$. In all cases, $w(0)$ transitions when $\Lambda(0)$ diverges.}
    \label{fig:phaseExample}
\end{figure}

We can test the behavior of $w(E)$ in multiple ways. First, if we consider the winding only around $E=0$ and set the onsite disorder $W=0$, then we can use the result of Ref. \onlinecite{mondragon2014topological} to analytically compute the localization length $\Lambda(0)$: 
\begin{equation}
    \frac{1}{\Lambda(0)} = \log\left(\frac{|J_R-\frac{W_R}{2}|^{\frac{J_R}{W_R}-\frac{1}{2}}|J_L+\frac{W_L}{2}|^{\frac{J_L}{W_L}+\frac{1}{2}}}{|J_R+\frac{W_R}{2}|^{\frac{J_R}{W_R}+\frac{1}{2}}|J_L-\frac{W_L}{2}|^{\frac{J_L}{W_L}-\frac{1}{2}}}\right).
\end{equation}
In Figs. \ref{fig:phaseExample}(a,b), we plot the numerically computed $w(0),$ and the analytically predicted curves where $\Lambda(0)=\infty,$ as a function of the disorder parameters $(W_L,W_R)$, at fixed model parameters (a) $(J_L,J_R,W)=(1,1,0)$ and (b) $(J_L,J_R,W)=(1,.5,0)$. We find that $w(0)$ is indeed quantized, and only changes when $\Lambda(0)$ diverges.
At more general $E$ and/or nonzero $W$, we can no longer analytically determine the phase boundaries. However, for a general point $(J_R,J_L,W)$ and general $E$, we can compute the localization length at $E$ numerically using transfer matrices\cite{pichard1981finite}. In Figs. \ref{fig:phaseExample}(c-f), we plot both $w(0)$ and  $\Lambda(0)$ for two systems with (c,e) $(J_L,J_R,W)=(1,1,1)$ and (d,f) $(J_L,J_R,W)=(1,.5,1)$. We again find that  $w(E)$ is quantized and transitions only when $\Lambda(E)$ diverges.

\begin{figure}
    \centering
    \includegraphics[width=\columnwidth]{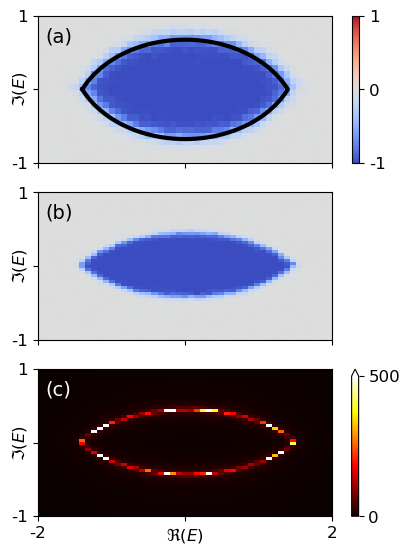}
    \caption{(a) $w(E)$ for a system with $W_L=W_R=0$ and $h^i$ distributed according to a Cauchy distribution. The black line denotes the values where $\Lambda(E)$ diverges. (b) $w(E)$ for a system with $(J_L,J_R,W_L,W_R,W)=(1.,5,1,1,1)$. (c) Numerically computed $\lambda(E)$. In each case, the region with $w(E)\neq 0$ is surrounded by a wall of delocalized states, and $w(E)$ only transitions when $\Lambda(E)$ diverges.}
    \label{fig:windingEnergy}
\end{figure}

To illustrate the localization properties of the spectrum we can also consider a system at fixed disorder, and calculate $w(E)$ as a function of $E$. If we assume $(W_L,W_R)=0$ and let $h^i$ follow a Cauchy distribution rather than the uniform distribution in Eq. \ref{eq:disorderParameters}, we can compute $\Lambda(E)$ analytically at any $E$\cite{thouless1972relation,goldsheid1998distribution}; this is shown in Fig. \ref{fig:windingEnergy}(a). For more general disorder configurations, we can again use the transfer matrix; an example with $(J_L,J_R,W_L,W_R,W)=(1.,5,1,1,1)$ is shown in Fig. \ref{fig:windingEnergy}(b,c). In both cases we find a region with non-vanishing winding completely surrounded by delocalized states. These examples reveal a notable distinction between the NH winding number $w(E)$ and Hermitian topological invariants such as the chiral winding number $\nu$ and the Chern number $C$. Indeed, $\nu$ is a stable topological invariant that is carried by entirely localized states\cite{mondragon2014topological}, while the Chern number can be carried by a single delocalized state in the spectrum\cite{halperin1982,prodan2010entanglement,onoda2007localization}. The NH winding, on the other hand, can be nonzero only when an extensive number of delocalized states exist in the spectrum. This is because $w(E)$ doesn't change under a continuous change of $E$ unless $E$ passes through a delocalized state, and $w(E)\rightarrow 0$ as $|E|\rightarrow\infty$ since in this limit $\hat Q$ approaches the identity matrix. This implies that any $E$ with a nonzero winding must be surrounded by a ``wall" of delocalized states, and the winding number prevents these states from localizing. Intuitively, an extensive sensitivity to boundary conditions should require an extensive number of delocalized states, and this is precisely what we find. 

\section{$w(E)$ and the NH skin effect}

In clean systems, we know a region having winding $w>0$ leads to an extensive number of states at the right edge of the system, and a region hiaving winding $w<0$ leads to an extensive number of states at the left edge of the system\cite{okuma2020topological,zhang2019correspondence}. We can characterize this NH skin effect by examining the density $\sum_n|\psi^n(x)|^2$ of all modes $\{\psi^n\}$ in the system. If there is a NH skin effect, the density at the corresponding edge should be proportional to the system size $N$, $\rho_\text{edge}=\Gamma N$. 

\begin{figure}
    \centering
    \includegraphics[width=\columnwidth]{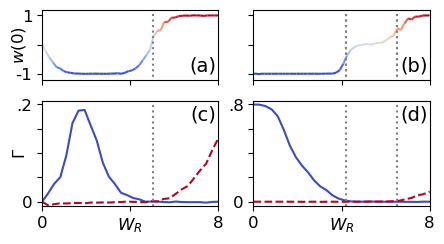}
    \caption{(a) $w(0)$ as a function of $W_R$ for $(J_L,J_R,W,W_L)=(1,1,0,0)$. This corresponds to the $x$-axis in Fig. \ref{fig:phaseExample}(a). (b) The same for $(J_L,J_R,W,W_L)=(1,.5,1,0)$, corresponding to the $x$-axis in Fig. \ref{fig:phaseExample}(e). (c,d) The corresponding slopes $\Gamma$ in the relation $\rho_{\text{edge}}=\Gamma N$ for the left (blue solid) and right (right dashed) edges. We see that the NH skin effect occurs at the left edge when $w<0$ and at the right edge when $w>0$.}
    \label{fig:nhskin}
\end{figure}

This behavior persists in the presence of disorder. For the Hatano-Nelson model, there is only one region of nonzero winding, centered at $E=0$, so $w(0)$ determines the NH skin effect. In Fig. \ref{fig:nhskin}(a,b), we plot the winding $w(0)$ as a function of $W_R$ for (a) $(J_L,J_R,W,W_L)=(1,1,0,0)$ and (b) $(J_L,J_R,W,W_L)=(1,.5,1,0)$. This corresponds to the $x$-axis of Figs. \ref{fig:phaseExample}(a) and (e). In Fig. \ref{fig:nhskin}(c,d) we plot the coefficient $\Gamma$ for each edge. Any $\Gamma>0$ indicates a NH skin effect. We see that, identical to the clean case, the NH skin effect occurs at the left edge when $w<0$ and at the right edge when $w>0$, and no skin effect occurs when $w=0$. More intricate models can have regions with both $w>0$ and $w<0$, and thus have a skin effect at both boundaries.

The connection between the NH skin effect and $w(E)$ allows us to predict a new phenomenon: the \emph{NH Anderson skin effect}, in which a system without a NH skin effect develops a skin effect at a critical value of disorder. Such an effect can already be seen in Fig. \ref{fig:phaseExample}c, in which the system near $(W_L,W_R)=(0,0)$ has $w(0)=0$ and thus no NH skin effect, but transitions to $w(0)=\pm 1$ at non-vanishing critical values of $W_L$ or $W_R$ (for example, by just tuning one of $W_L$ or $W_R$ and keeping the other zero). Such an effect should be readily observable in experimental platforms\cite{ghatak2019observation,brandenbourger2019non,zhou2020non,xiao2020non,hofmann2020reciprocal}.

As an immediate application, the connection between the NH skin effect and the winding number allows us to understand the stability of the recently demonstrated optical funnel based on the NH skin effect\cite{weidemann2020topological}. The optical funnel is a NH optical system in which all eigenmodes are localized at an interface; the effect of this localization is to ``funnel" all excitations (incident light) towards the interface. It was noted that for weak disorder numerics show weak funnelling still occurred, while at strong disorder it disappeared. Our formalism allows us to directly predict this phenomena by computing $w(E)$. In the clean limit the entire PBC spectrum surrounds a region with $w(E)\neq 0$, so that all OBC states are localized at the interface. As the disorder increases, the region with $w(E)\neq 0$ shrinks, so that the number of states localized at the interface decreases, leading to reduced funnelling. Finally, at a critical value of disorder, $w(E)=0$ everywhere and no funnelling occurs. Our formalism not only allows us to understand the stability of the funnelling to weak disorder, but provides a method to compute the critical value of disorder at which funnelling breaks down.

\section{Discussion}

In conclusion, we  extended the definition of the NH winding number $w(E)$ to disordered systems by relating it to the chiral winding number $\nu$ of a doubled Hermitian system. Our extension of $w(E)$ has several desirable properties. It is quantized, self-averaging, continuous as a function of parameters, and changes only when the localization length at $E$  diverges. In addition, our $w(E)$ successfully predicts the NH skin effect in disordered systems just as it does for clean systems. We have seen that, unlike Hermitian topological invariants, a nonzero $w(E)$ stabilizes an entire band of delocalized states in the region surrounding the nonzero $w(E)$. 

 Our prediction of a NH skin effect in strongly disordered systems, including the NH Anderson skin effect, should be experimentally verifiable, as there already exist multiple metamaterial platforms that exhibit the NH skin effect. In the future, it would be interesting to extend our results to higher dimensions, as the NH skin effect is not fully understood in higher dimensions even for clean systems\cite{lee2019hybrid,hofmann2020reciprocal,yoshida2020mirror}.

\section{Acknowledgements}

We thank Yuhao Ma for useful discussions. We thank Qi-Bo Zeng for helpful comments on an early version of the manuscript. We thank the US Office of Naval Research (ONR) Multidisciplinary University Research Initiative (MURI) grant N00014-20-1-2325 on Robust Photonic Materials with High-Order Topological Protection for support. We also thank the US National Science Foundation (NSF) Emerging Frontiers in Research and Innovation (EFRI) grant EFMA-1627184 for support. This work made use of the Illinois Campus Cluster, a computing resource that is operated by the Illinois Campus Cluster Program (ICCP) in conjunction with the National Center for Supercomputing Applications (NCSA) and which is supported by funds from the University of Illinois at Urbana-Champaign.

\bibliography{thebibliography.bib}

\onecolumngrid
\renewcommand{\thesection}{}
\renewcommand{\thesubsection}{S\arabic{subsection}}
\setcounter{table}{0}
\renewcommand{\thetable}{S\arabic{table}}
\setcounter{figure}{0}
\renewcommand{\thefigure}{S\arabic{figure}}
\setcounter{equation}{0}
\renewcommand{\theequation}{S\arabic{equation}}

\section*{Supplemental Material}

\subsection{Derivation and properties of the winding number $w(E)$}

In this section, we give a systematic derivation of our formula for the winding number $w(E)$ in disordered systems and prove it gives a robust topological invariant. For completeness, we reproduce the formula for the winding number in clean systems below

\begin{equation}
    w(E) = \frac{1}{2\pi i}\int_0^{2\pi} \partial_k \log\left( |\hat H_k-E|\right)\ dk
    \label{eq:Swinding}
\end{equation}

Following Ref. \onlinecite{gong2018topological}, we relate $w(E)$ to the topological invariant of a doubled Hermitian system $\hat{\mathcal{H}}$
\begin{equation}
    \hat{\mathcal{H}} = \left(\begin{matrix} 0 & \hat H -E\\\hat H^\dagger -E^* &0\end{matrix}\right)
    \label{eq:doubledH}
\end{equation}
By construction, $\hat{\mathcal{H}}$ has a chiral symmetry $\hat S=\left(\begin{matrix}1&0\\0&-1\end{matrix}\right)$ satisfying $\{\hat{S},\hat{\mathcal{H}}\}=0$. Translationally invariant 1D systems with a chiral symmetry are classified by a chiral-symmetric winding number $\nu$ defined by
\begin{equation}
    \nu = \frac{1}{2\pi i}\int_0^{2\pi} \partial_k \log\left(|\hat S^+\hat{\mathcal{H}}_k \hat{\mathcal S}^-|\right)\ dk
    \label{eq:windingChiral}
\end{equation}
where $\hat{\mathcal S}^\pm$ is the projector onto the $\pm 1$ subspace of $\hat{S}$. We see that $\hat{\mathcal S}^+\hat{\mathcal{H}}\hat{\mathcal S}^-=\hat H_k-E$, thus $\nu=w(E)$.

The chiral winding number $\nu$ can be generalized to disordered chiral-symmetric systems in terms of the projector $\hat{\mathcal{P}}$ onto the eigenstates with energy $<0$\cite{mondragon2014topological,song2014aiii,prodan2016non}:
\begin{equation}
    \nu=\mathcal{T}\left\{\hat{\mathcal Q}^{-+}[\hat{\mathcal Q}^{+-},\hat{\mathcal X}]\right\},\qquad \hat{Q}\equiv \mathbbm{1}-2\hat{\mathcal P},\qquad \hat{\mathcal Q}^{\pm\mp} = \hat{\mathcal S}^\pm\hat{\mathcal Q}\hat{\mathcal S}^\mp.
    \label{eq:defineNu}
\end{equation}
Refs. \onlinecite{song2014aiii} and \onlinecite{prodan2016non} proved that $\nu$ is quantized, self-averaging, and continuous with respect to parameters in $\hat{\mathcal{H}}$, provided no delocalized eigenstate exists at $E=0$. Therefore, it is a topological invariant that cannot change without a mobility gap closing. Physically, $\nu$ predicts the number of zero-energy chiral edge modes in a half-infinte system with boundary\cite{prodan2016non,prodan2016bulk}. If the half-infinite system extends to $-\infty$, there are $|\nu|$ protected zero-energy edge modes with chirality $-\text{sgn}(\nu)$, while if the half-infinite system extends to $+\infty$ there are $|\nu|$ protected zero-energy edge modes with chirality $\text{sgn}(\nu)$.

Generalizing the equality between $\nu$ and $w(E)$ for clean systems, we postulate that the proper generalization of $w(E)$ to systems with disorder is simply $w(E)=\nu$. As written, the formula for $w(E)$ refers explicitly to an artificial doubled Hamiltonian $\hat{\mathcal H}$, and the properties of $w(E)$ are functions of $\hat{\mathcal H}$. In the following, we remedy this by giving an explicit formula for $w(E)$ in terms of the original Hamiltonian $\hat{H}$, and translate the properties of $w(E)$ to refer only to the original Hamiltonian $\hat{H}$.

First, we derive how to calculate $w(E)$ quantity directly from $\hat{H}$ and $E$ without reference to a doubled Hamiltonian. This can be achieved using the polar decomposition $(\hat{H}-E)=\hat{Q}\hat{P}$, where $\hat{Q}$ is unitary and $\hat{P}$ is a positive Hermitian matrix. We can then write Eq. \ref{eq:doubledH} as
\begin{equation}
    \hat{\mathcal{H}} = \left(\begin{matrix} 0 & \hat Q\hat P\\\hat P\hat Q^\dagger &0\end{matrix}\right).
\end{equation}
If $|\psi_i\rangle$ is a complete basis of eigenstates of $\hat{P}$ with eigenvalues $\lambda_i$, then it is simple to show that the eigenvectors of $\hat{\mathcal H}$ are $|\Psi_i^\pm\rangle=\frac{1}{\sqrt 2}\left(\begin{matrix} \pm{\hat Q}|\psi_i\rangle \\|\psi_i\rangle\end{matrix}\right)$ with eigenvalues $\lambda_i^\pm$. Then (Eq. \ref{eq:defineNu}) $\hat{\mathcal Q}$ can be written as
\begin{equation}
    \hat{\mathcal Q} = \mathbbm{1}-2\sum_i|\Psi_i^-\rangle\langle\Psi_i^-|=\mathbbm{1}-\sum_i\left(\begin{matrix} \hat{Q}|\psi_i\rangle\langle\psi_i|\hat Q^\dagger & -\hat Q|\psi_i\rangle\langle\psi_i| \\-|\psi_i\rangle\langle\psi_i|\hat Q^\dagger & |\psi_i\rangle\langle\psi_i|\end{matrix}\right)=\left(\begin{matrix} 0 & \hat Q \\Q^\dagger & 0\end{matrix}\right)
\end{equation}
Thus, we see that $\hat{\mathcal{Q}}^{+-}=\hat Q$, and $\hat{\mathcal{Q}}^{-+}=\hat Q^\dagger$. Thus, our final formula for $w(E)$ is given by
\begin{equation}
    w(E) = \mathcal{T}\left\{\hat{Q}^\dagger[\hat Q,\hat X]\right\}
    \label{eq:disorderWinding}
\end{equation}

Next, we consider the quantization, self-averaging, and continuity of $w(E)$. We know that $\nu$ is quantized, self-averaging, and continuous with respect to parameters of $\hat{\mathcal H}$. From Eq. \ref{eq:doubledH} for $\hat{\mathcal{H}}$, we see that this immediately implies that $w(E)$ is quantized, self-averaging, and continuous with respect to both $E$ and parameters of $\hat H$.

We can also get a physical interpretation of $w(E)$ for our original Hamiltonian $\hat{H}$. If $w(E)>0$, then we know that $\hat{\mathcal H}$ has $|w(E)|$ zero-energy edge modes of negative chirality for a half-infinite system that extends to $-\infty$. Explicitly, we have
\begin{equation}
    \left(\begin{matrix} 0 & \hat{H}-E\\\hat{H}^\dagger-E^* &0\end{matrix}\right)\left(\begin{matrix}0\\|\psi\rangle\end{matrix}\right)=\left(\begin{matrix}(\hat{H}-E)|\psi\rangle\\0\end{matrix}\right)=\left(\begin{matrix}0\\0\end{matrix}\right)
\end{equation}
which says that $|\psi\rangle$ is an edge eigenstate with energy $E$ in the half-infinite system. Similarly, if $w(E)<0$, then we know that $\hat{\mathcal H}$ has $|w(E)|$ zero-energy edge modes of negative chirality for a half-infinite system that extends to $+\infty$. We again find that if we write $\left(\begin{matrix}0\\|\psi\rangle\end{matrix}\right)$ for the zero-energy edge mode of $\hat{\mathcal{H}}$, then $|\psi\rangle$ is an edge eigenstate with energy $E$. In total, we find that \emph{positive} winding numbers predict protected edge-localized eigenstates on a semi-infinite system extending to $-\infty$, and \emph{negative} winding numbers predict protected edge-localized eigenstates on a semi-infinite system extending to $+\infty$. The connection between $w(E)$ and the presence of protected edge-localized eigenstates justifies our Eq. \ref{eq:disorderWinding} as the proper generalization of the winding number to disordered systems. Note that our discussion has focused on \emph{right} eigenstates; the same logic predicts protected \emph{left} edge eigenstates at energy $E$ for $w(E)$ of the opposite sign.

Finally, we consider when $w(E)$ can transition. We know that $\nu$ can only change when $\hat{\mathcal{H}}$ has a delocalized eigenstate at $E=0$. Writing this condition out as 
\begin{equation}
    \left(\begin{matrix} 0 & \hat{H}-E\\\hat{H}^\dagger-E^* &0\end{matrix}\right)\left(\begin{matrix}|\psi_L\rangle\\|\psi_R\rangle\end{matrix}\right)=\left(\begin{matrix}(\hat{H}-E)|\psi_R\rangle\\(\hat{H}^\dagger-E^*)|\psi_L\rangle\end{matrix}\right)=\left(\begin{matrix}0\\0\end{matrix}\right)
\end{equation}
we see that $w(E)$ can only change if $\hat H$ has a delocalized left eigenstate ($\langle\psi_L|$) \emph{or} right eigenstate ($|\psi_R\rangle$) with eigenvalue $E$. However, any one-band NH tight-binding model has a symmetry that relates the localization of left eigenstates and right eigenstates. We can write a general one-band Hamiltonian as
\begin{equation}
\hat{H} = \sum_i\sum_r \hat{c}_{(i+r)}^\dagger {h}_i^r  \hat{c}_{i}
\end{equation}
where and ${h}_i^r$ is an $i$-dependent random hopping connecting site $i$ to site $(i+r)$ drawn from a random distribution $P_r$ independent of $i$. Then $\hat{H}^\dagger$ is given by
\begin{equation}
\hat{H} = \sum_i\sum_r \hat{c}_{i}^\dagger ({h}_i^r)^*  \hat{c}_{(i+r)}
\end{equation}
If we define the mirror symmetry $\hat{M}$ by $\hat{M}\hat{c}_{i}\hat{M}=\hat{c}_{(-i)}$, then 
\begin{equation}
\hat{M}\hat{H}^\dagger\hat{M} = \sum_i\sum_r \hat{c}_{(-i)}^\dagger ({h}_i^r)^*  \hat{c}_{(-i-r)}=\sum_i\sum_r \hat{c}_{(i+r)}^\dagger ({h}_{(-i-r)}^r)^*  \hat{c}_{i}
\end{equation}
This shows that $\hat{M}\hat{H}^\dagger\hat{M}$ is unitarily equivalent to the complex conjugate of a matrix that is drawn from the same disorder distribution $\{P_r\}$ as $\hat{H}$. Thus, if $|\psi_R\rangle$ is a right eigenstate of $\hat{H}$ with eigenvalue $E$, then $\hat{M}|\psi_R^*\rangle$ is an eigenstate of $(\hat{H'})^\dagger$ with eigenvalue $E^*$, where $\hat{H'}$ is a Hamiltonian from the same disorder distribution as $\hat{H}$. Since the localization length is a property of the distribution of Hamiltonians, not the Hamiltonian itself, 
this shows that $\hat{H}$ has a delocalized right eigenstate with eigenvalue $E$ if and only if it has a delocalized left eigenstate with eigenvalue $E$. Therefore, we can simply say that $w(E)$ only changes when $\hat{H}$ has a delocalized state at $E$. For general multiband models, we cannot eliminate the possibility that the localization lengths of the left and right eigenvectors are not equal, although this seems unlikely.

\end{document}